\documentstyle[12pt]{article}

\newcommand{\chose}[2]{\bigg(\hspace{-1mm}
\begin{array}{c}{\scriptstyle #1}\\
{\scriptstyle #2}\end{array}\hspace{-1mm}\bigg)}

\begin{document}

\begin{center}

\bigskip
{\Large Between entropy and subentropy}\\

\bigskip

Sarah R.~Nichols and William K.~Wootters\\

\bigskip

{\small{\sl

Department of Physics, Williams College, Williamstown, 
MA 01267, USA }}\vspace{3cm}

\end{center}
\subsection*{\centering Abstract}
{The von Neumann entropy and the subentropy of
a mixed quantum state are upper and lower bounds,
respectively, on the accessible information of
any ensemble consistent with the given mixed
state.  Here we define and investigate a set
of quantities intermediate between entropy
and subentropy.}

\vfill

PACS numbers: 03.67.-a, 89.70.+c, 65.40.Gr

\newpage

\section{Entropy and subentropy} 
The von Neumann entropy of a quantum 
state $\rho$ can be defined as 
\begin{equation}
S(\rho) = -\hbox{tr}\,\rho\ln\rho =
-\sum_{j=1}^n \lambda_j \ln \lambda_j, \label{S}
\end{equation}
where $n$ is the dimension of $\rho$,
the $\lambda$'s are its eigenvalues, 
and the expression
$x\ln x$, when evaluated at $x=0$, is taken to have 
the value $\lim_{x\rightarrow 0}x\ln x = 0$.  
The von Neumann entropy is
of central importance in physics; 
when applied to a thermal ensemble, it is
{\em the} entropy of thermodynamics.
In quantum information theory it plays prominent
roles in many contexts, {\em e.g.}, in studies of the 
classical capacity of a quantum channel \cite{SW,H}
and 
the compressibility of a quantum source \cite{S,SJ}.
To introduce the problem 
that we will be considering here, we focus
on the role that the von Neumann entropy plays
in Holevo's theorem \cite{Hthm,FC,Yuen,SWW}.
Part of the content of
this theorem can be stated as follows.  
Suppose we are handed a quantum object
and are told that it is in one of several possible pure states
$|\psi_i\rangle$, $i=1,\ldots, N$, the probability
of the state $|\psi_i\rangle$ being $p_i$.  
By measuring this single object, we aim
to get as much information as possible 
about the identity of the state, that is, 
the value of the index $i$.
The maximum amount we can obtain is called the 
{\em accessible 
information} of the ensemble consisting of
the ordered pairs $(|\psi_i\rangle,p_i)$.
In general there is no analytic formula
for the accessible information, but Holevo's 
theorem gives us a 
simple and 
general upper bound: the accessible information
is no greater than the von Neumann entropy of
the ensemble's density matrix
\begin{equation}
\rho = \sum_{i=1}^N p_i |\psi_i\rangle\langle\psi_i|.
\end{equation}
Moreover, 
the von Neumann entropy---we will usually refer 
to it simply as the entropy---is the 
{\em least} upper bound on the accessible information
that depends only on the density matrix $\rho$
and not on other details of the ensemble.
To see why this is true, note that the ensemble consisting
of the eigenstates of $\rho$, with the eigenvalues
as weights, is an ensemble realizing the density matrix $\rho$
and from which one can extract, in a single measurement,
an amount of information
equal to $S(\rho)$.  That is, the upper bound can
be achieved.

It is natural also to ask about the analogous
{\em lower} bound: what is the greatest
lower bound on the accessible information of an
ensemble that depends only on the ensemble's
density matrix?  This question has been answered \cite{JRW}: 
the greatest lower bound is
the {\em subentropy} $Q(\rho)$, defined by
\begin{equation}
Q(\rho) = -\sum_{j=1}^n \Bigg(\prod_{k\neq j}
\frac{\lambda_j}{\lambda_j-\lambda_k}
\Bigg) \lambda_j \ln \lambda_j.  \label{Q}
\end{equation}
(If two or more of the eigenvalues $\lambda_j$
are equal, the value of $Q$ is determined unambiguously
by taking a limit starting with unequal eigenvalues.)
Just as the ensemble of eigenstates of $\rho$ 
has an accessible information that matches the
upper bound $S(\rho)$, there is a complementary
ensemble, called the Scrooge ensemble \cite{JRW}, that likewise
realizes $\rho$ but has an accessible information
equal to the lower bound $Q(\rho)$. 

Thus in this context of acquiring information
from a single quantum system, the von Neumann
entropy and its lesser known analog the subentropy
play mirror-image roles and together define the
range of possible values of the accessible information 
for a given density matrix.

Comparing Eqs.~(\ref{S}) and (\ref{Q}) one sees
a certain formal similarity between $S$ and $Q$.  
The similarity is more striking if we rewrite
both $S$ and $Q$ as contour integrals \cite{JRW}.  One
can write 
\begin{equation}
S(\rho) = -\frac{1}{2\pi i}
\oint (\ln z)\,\hbox{tr}\,(I-\rho /z)^{-1}dz,
\label{Sint}
\end{equation}
where the contour encloses all the nonzero
eigenvalues of $\rho$.  To make the connection
between Eq.~(\ref{Sint}) and Eq.~(\ref{S})
note that the eigenvalues of \hbox{$(I-\rho/z)^{-1}$}
are $z/(z-\lambda_j)$, so that each term in the
trace contributes a residue that becomes a term
in Eq.~(\ref{S}).
Similarly, one can express $Q$ as
\begin{equation}
Q(\rho) = -\frac{1}{2\pi i}
\oint (\ln z)\,\hbox{det}\,(I-\rho /z)^{-1}dz.
\label{Qint}
\end{equation}
Thus, where the trace appears in the formula
for entropy, the determinant appears in the 
formula for subentropy.

The formulas given in Eqs.~(\ref{Sint}) and
(\ref{Qint}) raise an interesting mathematical
issue which is the impetus for this paper.
The trace and the determinant of a matrix are
simply the first and last of the coefficients
in the characteristic polynomial of the 
matrix.  In place of the trace in Eq.~(\ref{Sint})
or the determinant in Eq.~(\ref{Qint}),
one could insert any of the other coefficients
of this polynomial and thereby identify new
functions that might be regarded as natural generalizations
of entropy and subentropy.  In what follows we
define a set of functions of $\rho$ based
on this mathematical substitution and
investigate their properties.   
We call the functions $R^{(n)}_r$, $r=
1,\ldots,n$, with $R^{(n)}_1$ being equal
to $S$ and
$R^{(n)}_n$ being equal to $Q$.  
Among the properties
we will discover is the string of
inequalities
$Q = R^{(n)}_n\leq R^{(n)}_{n-1} \leq \cdots
\leq R^{(n)}_1 = S$, valid for any density
matrix $\rho$.  

In some respects, the subentropy $Q$ is quite unlike
the entropy $S$.  For example, $Q$ is not additive:
if $\rho = \rho_1\otimes\rho_2$, then $Q(\rho)$ is typically
not the same as $Q(\rho_1)+Q(\rho_2)$, whereas the
entropy is always additive in this sense.  However, $Q$
does share with $S$ the following property.  Suppose
we augment the state space and the density matrix
$\rho$ by including $m$ 
extra dimensions with zero weight.  That is,
we replace $\rho$ with $\rho \oplus 0_{m}$, where
$0_m$ is the $m\times m$ zero matrix, in effect adding
to the set of eigenvalues $(\lambda_1,\ldots,\lambda_n)$
$m$ additional eigenvalues all equal to zero.  One can
see immediately from Eqs.~(\ref{S}) and (\ref{Q}) that
both $S$ and $Q$ remain invariant under this
augmentation of the space.  Since we are looking
for natural generalizations of $S$ and $Q$, 
it is interesting to
ask whether our new quantities $R^{(n)}_r$ also have this
property.  

We will find, in fact, that they do not.
But we will be able to construct simple 
convex combinations
of the $R^{(n)}_r$'s that do remain invariant under
the addition of ``null'' dimensions.  These particular
linear combinations, called ${\mathcal R}_\alpha$,
are parameterized by the single continuous
parameter $\alpha$ and interpolate between $S$ and $Q$.

We are thus investigating in this paper various functions that
generalize von Neumann entropy and subentropy
in a specific mathematical sense.  There is no
guarantee, of course, that these functions will
be of value for physics.  At the
end of the paper we offer a speculative potential 
interpretation of ${\mathcal R}_\alpha$ in
quantum information theory
but otherwise leave this question
for future investigation.  

\section{Definition of $R^{(n)}_r$}
Given any $n\times n$ complex matrix $M$, the 
characteristic polynomial of $M$ is the quantity
$\det(\mu I - M)$ regarded as a function of 
$\mu$.  If we write this polynomial as
\begin{equation}
\det(\mu I - M) = \mu^{n} + \sum_{r=1}^{n} (-1)^{r}C_r(M) \mu^{n-r},
\end{equation}
then the coefficient $C_r(M)$ is given by
\begin{equation}
C_r(M) = \sum_{k_1<\cdots < k_r} 
\bigg(\prod_{s=1}^r \nu_{k_s}\bigg),
\end{equation}
the $\nu$'s being the eigenvalues of $M$.
Thus the index $r$ indicates the number of 
eigenvalues being multiplied together in each
term.\footnote{We
adopt the convention that there are always exactly
$n$ eigenvalues of an $n\times n$ matrix: if a root 
$\mu=\nu$
of the equation
$\det(\mu I - M)=0$ has multiplicity $m$, we say that
$m$ of the $n$ eigenvalues of $M$ have the value $\nu$.}  The coefficient $C_1(M)$ is the trace of $M$,
and $C_n(M)$ is the determinant.  

By analogy with Eqs.~(\ref{Sint}) and (\ref{Qint}),
we now define a set of quantities $R^{(n)}_r$
as follows:
\begin{equation}
R^{(n)}_r(\rho) = -{\chose{n-1}{r-1}}^{-1}
\frac{1}{2\pi i}
\oint (\ln z)C_r[(I-\rho/z)^{-1}] dz, \label{Rint}
\end{equation}
where again the contour is chosen to enclose all
the nonzero eigenvalues of $\rho$, and 
$\chose{n-1}{r-1}$ is the binomial coefficient
$\frac{(n-1)!}{(r-1)!(n-r)!}$.  We have included
this factor because, as we will see in the 
following sections, it places the functions $R^{(n)}_r$
between $S$ and $Q$.  Note that, as promised,
$R^{(n)}_r(\rho)$ is equal to $S(\rho)$ for
$r=1$ and to $Q(\rho)$ for $r=n$.  

It is straightforward to evaluate the integral in
Eq.~(\ref{Rint}) so as to write $R^{(n)}_r$ explicitly
as a function of the eigenvalues of $\rho$.
One finds that 
\begin{equation}
R^{(n)}_r = -\chose{n-1}{r-1}^{-1}\hspace{-2mm}
\sum_{k_1<\cdots <k_r}\sum_{s=1}^r
\Bigg[
\prod_{t\neq s}^r \bigg(\frac{\lambda_{k_s}}
{\lambda_{k_s}-\lambda_{k_t}}\bigg)\Bigg]
\lambda_{k_s}\ln\lambda_{k_s}.
\label{R1}
\end{equation}
For $r=2, \ldots, n$, we can rearrange the
indices to get an expression more analogous
to Eq.~(\ref{Q}):
\begin{equation}
R^{(n)}_r = -\sum_{j=1}^n\Bigg\{
\chose{n-1}{r-1}^{-1}\hspace{-5mm}
\sum_{\begin{array}{c}
{\scriptstyle k_1<\cdots < k_{r-1}}\\ \hbox{\small each}\,\,
{\scriptstyle k_s
\neq j} \end{array}}\hspace{-2mm}
\Bigg[\prod_{s=1}^{r-1}\bigg(\frac{\lambda_j}
{\lambda_j - \lambda_{k_s}}\bigg)\Bigg]
\Bigg\}\lambda_j\ln\lambda_j.
\label{R}
\end{equation}
Notice that the number of terms in the
sum over $k_1, \ldots, k_{r-1}$ is 
$\chose{n-1}{r-1}$, because there are
$n-1$ index-values from which to choose,
the value
$j$ being disallowed.  Thus the quantity
in curly brackets is an {\em average} of the kind
of product that appears in the expression
(\ref{Q}) for $Q$.

As in the case of $Q$, in order to evaluate
Eq.~(\ref{R}) when two or more of the
eigenvalues $\lambda_j$ are equal, we 
have to take a limit.  That the limit is
unique is guaranteed by Eq.~(\ref{Rint})
which has a unique value for all density
matrices $\rho$.  

Though we have already written the
functions $R^{(n)}_r$
in a few ways, it will be helpful
to re-express these functions in quite different
terms in order to derive certain properties.
This re-expression is the goal of the following
section.

\section{Another path to $R^{(n)}_r$}
Let us return to the problem of ascertaining
the quantum state of a single quantum system,
given the ensemble $\{(|\psi_i\rangle,p_i)\}$.
In addition to being a lower
bound on the amount of information one
can gain when one makes the {\em best}
possible measurement, the 
subentropy $Q(\rho)$ is also the 
{\em average} information one obtains
about the state, where the average is
over all complete orthogonal measurements.
(Indeed, the latter fact is sufficient to prove
that $Q$ is a lower bound on the accessible
information.)  Interpreting $Q$ as this average
leads to another way of expressing $Q$
mathematically \cite{JRW}.  
\begin{equation}
Q(\rho) = -n\int \bigg(\sum_i\lambda_i x_i\bigg)
\ln\bigg(\sum_i\lambda_i x_i\bigg)dx
+n\int x_1 \ln x_1 dx.
\label{Qave}
\end{equation}
Here the $x_i$'s are non-negative real numbers
constrained to sum to unity; that is, the ordered
set $x=(x_1,\ldots,x_n)$ represents a point in
the probability space, or probability simplex, 
appropriate for a set of
$n$ possibilities.  The integrals in Eq.~(\ref{Qave})
are integrals over this probability
space, the measure being
the uniform measure normalized
to unity.
Explicitly, for any function $g(x)$,
\begin{equation}
\int g(x) dx \equiv \frac{1}{(n-1)!}
\int_0^1 \int_0^{1-x_1}\cdots\int_0^{1-x_1-\cdots -x_{n-2}}
g(x) dx_{n-1}\cdots dx_2 dx_1.
\end{equation}
In Eq.~(\ref{Qave}) there is no special significance
to the index 1 that appears in the second integral.
Because of the symmetry of the measure, any other
of the $x_i$'s could equally well have been chosen.
In fact, we can write the integral more symmetrically
as follows.
\begin{equation}
Q(\rho) = n\int f(x) dx,
\label{Qf}
\end{equation}
where 
\begin{equation}
f(x) = -\bigg(\sum_i\lambda_i x_i\bigg)
\ln\bigg(\sum_i\lambda_i x_i\bigg)
+ \sum_i \lambda_i x_i \ln x_i.
\label{f}
\end{equation}

Interestingly, the entropy $S(\rho)=R^{(n)}_1(\rho)$ can be written 
in an analogous form.
We simply need to replace the integral $\int(\cdots)dx$
in Eq.~(\ref{Qf})
with a discrete sum over the extreme points of the 
probability simplex.  That is, instead of integrating
over all points $x = (x_1, \ldots, x_n)$, we sum over
the special points $x^{(1)}=(1,0,\ldots, 0)$, 
$x^{(2)}=(0,1,0,\ldots,0)$,
\ldots, $x^{(n)}=(0,\ldots, 0,1)$.  Again, we take the
total weight
of all these points to be unity.  Thus, starting
with Eq.~(\ref{Qf}) we perform the
modification
\begin{equation}
\int f(x) dx \rightarrow \frac{1}{n}\sum_{j=1}^n f(x^{(j)}),
\end{equation}
which brings us to 
\begin{equation}
n\bigg(\frac{1}{n}\bigg)\sum_{j=1}^n f(x^{(j)}) = 
-\sum_{j=1}^n \lambda_j \ln \lambda_j = S(\rho).
\end{equation}

It turns out that the quantities $R^{(n)}_r$ for other
values of $r$ can likewise
be expressed as in Eq.~(\ref{Qf}) but with different
ranges of integration.  We have just seen that $R^{(n)}_1$,
which is the entropy itself, can be expressed in this 
way if the ``integral'' is taken to be over the discrete
set of extreme points of the simplex.  As we will show 
shortly, $R^{(n)}_2$
is similarly given by Eq.~(\ref{Qf}), but with the integral
being taken
over the {\em edges} of the simplex, that is, over those
points $x$ having at most two nonzero components.  (Again
the measure is uniform in the Euclidean sense and normalized
to unity.)  And in general, $R^{(n)}_r$ is given by the
same expression, but with the integral being over all points
$x$ having at most $r$ nonzero components.

To prove this claim, let us set up the integral $I^{(n)}_r$ 
that we have just described:
\begin{equation}
I^{(n)}_r = \chose{n}{r}^{-1}n\sum_{k_1<\cdots <k_r}
\int_{k_1,\ldots,k_r} f(x) dx.
\label{Ir}
\end{equation}
Here $\int_{k_1,\ldots,k_r} (\cdots) dx$ is the integral
over the ``face'' of the simplex in which only
$x_{k_1}, \ldots, x_{k_r}$ are nonzero, with the 
measure normalized 
to unity.  There are $\chose{n}{r}$ terms in the sum,
so we have divided by $\chose{n}{r}$ to ensure
that the measure of the entire region over which
we are integrating---that is, the collection of all the relevant 
faces---is normalized to unity.
We wish to show that $I^{(n)}_r = R^{(n)}_r$.
 
Consider first the integral over just one face,
\begin{equation}
\int_{k_1,\ldots,k_r} f(x) dx.
\end{equation}
We can regard this integral as being over a complete
probability space, but with only $r$ possibilities
instead of $n$.  Therefore, if we multiply it by $r$,
we see from Eq.~(\ref{Qf}) that we get something
formally similar to $Q$---not the $Q$ of the original 
density matrix $\rho$
but rather of an effective $r$-dimensional
density matrix whose (unnormalized) eigenvalues
are $\lambda_{k_1}, \ldots, \lambda_{k_r}$.  (The
equivalence between Eq.~(\ref{Qf}) and 
Eq.~(\ref{Q}) does not depend on the 
$\lambda$'s adding up to unity \cite{JRW}.)  That is, from
Eq.~(\ref{Q}) we have
\begin{equation}
\int_{k_1,\ldots,k_r} f(x) dx =-\bigg(\frac{1}{r}\bigg)
\sum_{s=1}^r
\Bigg[
\prod_{t\neq s}^r \bigg(\frac{\lambda_{k_s}}
{\lambda_{k_s}-\lambda_{k_t}}\bigg)\Bigg]
\lambda_{k_s}\ln\lambda_{k_s}.
\end{equation}
Inserting this expression into Eq.~(\ref{Ir}),
we get
\begin{equation}
I^{(n)}_r = -\chose{n-1}{r-1}^{-1}\hspace{-2mm}
\sum_{k_1<\cdots <k_r}\sum_{s=1}^r
\Bigg[
\prod_{t\neq s}^r \bigg(\frac{\lambda_{k_s}}
{\lambda_{k_s}-\lambda_{k_t}}\bigg)\Bigg]
\lambda_{k_s}\ln\lambda_{k_s},
\label{I=R}
\end{equation}
which according to Eq.~(\ref{R1}) is equal
to $R^{(n)}_r$.  We have, therefore,
\begin{equation}
R^{(n)}_r = I^{(n)}_r = 
\chose{n}{r}^{-1}n\sum_{k_1<\cdots <k_r}
\int_{k_1,\ldots,k_r} f(x) dx,
\label{Rf}
\end{equation}
as claimed.

We can thus write all the quantities
$R^{(n)}_r$ as normalized integrals of the same
integrand, but with different ranges
of integration.

\section{Ordering the $R$'s}
In this section we use the form just derived
to prove the string of inequalities
mentioned in the introduction:
\begin{equation}
Q(\rho) = R^{(n)}_n(\rho)
 \leq R^{(n)}_{n-1}(\rho) \leq \cdots \leq R^{(n)}_1(\rho)
 = S(\rho),
\label{ineq}
\end{equation}
which hold for every $n\times n$
density matrix $\rho$.  We will show, in fact,
that all the inequalities are {\em strict} except when
$\rho$ is pure, in which case $R^{(n)}_r=0$ for every
$r$.
Since each function $R^{(n)}_r$ 
depends only on the eigenvalues $\lambda_1,\ldots,
\lambda_n$, which are non-negative and sum to unity, we 
can alternatively think of $R^{(n)}_r$ as a 
function on the probability space for a set
of $n$ possibilities.  If we picture each of these functions
as a ``surface'' plotted over the probability 
space, our inequalities tell us that
the surfaces corresponding to different values
of $r$ do not cross each other and coincide only
at the extreme points of the simplex.

To prove the (non-strict) inequalities (\ref{ineq}), we first
prove that the function $f$ defined in Eq.~(\ref{f})
is a convex function of $x$ for every set of allowed values
of the $\lambda$'s.  We do this by extending the definition
(\ref{f}) to all non-negative values of the $x_i$'s---that
is, we allow $x$ to be unnormalized---and showing that
$f$ is convex even in this larger set.  Treating
the $x_i$'s as independent variables---and for the 
moment restricting our attention to the case where
they are all strictly positive---let us compute
the matrix of second derivatives of $f$:
\begin{equation}
M_{ij} \equiv \Big(\sum_k \lambda_k x_k\Big)
\frac{\partial^2 f}{\partial x_i \partial x_j}
=  \delta_{ij}
\frac{\lambda_i}{x_i}\Big(\sum_k \lambda_k x_k\Big)
-\lambda_i\lambda_j.
\end{equation}
We show that the matrix $M$ is non-negative definite
by considering its expectation value with respect 
to an arbitrary real vector $v$. Using Dirac notation, 
we have 
\begin{equation}
\langle v|M |v\rangle = 
\bigg(\sum_i \frac{v_i^2 \lambda_i}{x_i}
\bigg)\Big(\sum_k \lambda_k x_k\Big)
-\Big(\sum_i v_i \lambda_i\Big)^2.
\end{equation}
But if we define new vectors $w$ and $z$ by 
$w_i = v_i \sqrt{\lambda_i/x_i}$ and
$z_i = \sqrt{\lambda_i x_i}$, then we can
write this equation as
\begin{equation}
\langle v|M|v\rangle = \langle w|w\rangle
\langle z|z\rangle - \langle w|z\rangle^2,
\label{Schwartz}
\end{equation}
whose right-hand side is non-negative by the 
Schwartz inequality.
Because $M$ is related to 
$\partial^2 f/\partial x_i \partial x_j$ by a 
positive factor, it follows that 
$f$ is a convex function of $x$, at least 
when each $x_i$ is greater than zero.  But by continuity, the
convexity extends to those points where some
of the
components $x_i$ are zero.

We will also need {\em strict} convexity in certain cases,
and for this we need to take into account the possibility
that some of the $\lambda$'s might be zero.  Suppose that 
$\lambda_{k_1},\ldots,\lambda_{k_s}$ are nonzero and that
all the other $\lambda$'s are zero.
Notice that in that case the right-hand side of Eq.~(\ref{Schwartz})
is zero only when the components $(v_{k_1},\ldots,
v_{k_s})$ of $v$ are proportional to the corresponding
components $(x_{k_1},\ldots,x_{k_s})$ of $x$.  But $v$
defines the direction along which we are taking the 
second derivative of $f$.    
Therefore if we consider a line containing
two values of $(x_{k_1},\ldots,x_{k_s})$ 
that are {\em not} proportional to
each other, the second
derivative of $f$ along this line is {\em strictly} positive, 
so that $f$ is strictly convex along this line.  (The
second derivative might approach infinity as some components
$x_i$ approach zero, but this pathology 
does not ruin the convexity.)
We will need this fact shortly.

We now use the convexity of $f$ to prove the
inequalities (\ref{ineq}), beginning with
the first one:
$R^{(n)}_{n} \leq R^{(n)}_{n-1}$.
Consider any point $x=(x_1,\ldots,x_n)$
in the probability simplex that is not one
of the extreme points.
We can write $x$ as
\begin{eqnarray}
&(x_1,\ldots,x_n) = \frac{1}{n-1}
\Big\{(1-x_1)[(0,x_2,\ldots,x_n)/(1-x_1)]\nonumber \\
&+(1-x_2)[(x_1,0,x_3,\ldots,x_n)/(1-x_2)] 
+ \cdots \label{long} \\
&+(1-x_n)[(x_1,\ldots,x_{n-1},0)/(1-x_n)]\Big\}.\nonumber
\end{eqnarray}
Notice that the vectors in square brackets are
all properly normalized, and that the coefficients
multiplying them, that is, $\,(1\, -\, x_1)/(n\, -\, 1)
\, ,\, \ldots\, ,$ 
\hbox{$(1-x_n)/(n-1)$}, add up to one.  We have thus written
the vector $x$ as an average of other legitimate
probability vectors.  
From the convexity of $f$, it follows then that
\begin{eqnarray}
&f(x) \leq \frac{1}{n-1}
\Big\{(1-x_1)f[(0,x_2,\ldots,x_n)/(1-x_1)]\nonumber \\
&+(1-x_2)f[(x_1,0,x_3,\ldots,x_n)/(1-x_2)] 
+ \cdots \label{flong} \\
&+(1-x_n)f[(x_1,\ldots,x_{n-1},0)/(1-x_n)]\Big\}.\nonumber
\end{eqnarray}
Moreover, if any two of the $\lambda_i$'s are nonzero, and
if the corresponding components $x_i$ are also nonzero (we
are about to integrate over all $x$, so that this latter
condition is almost always met), then
for at least one pair of the normalized
vectors appearing in
Eq.~(\ref{long}), the line connecting them is a line
along which $f$ is {\em strictly} convex.  Thus in this
case the
inequality in Eq.~(\ref{flong}) is strict. 

We now integrate both sides of the inequality (\ref{flong})
over the whole probability simplex, again using our
normalized measure.  
To see what this integration does to the right-hand
side, let us consider for now just the first term,
\begin{equation}
\frac{1}{n-1}\int 
(1-x_1)f[(0,x_2,\ldots,x_n)/(1-x_1)] dx.
\label{term1}
\end{equation}
We perform the integral by first integrating
over each surface that has a fixed value of $x_1$,
and then integrating over $x_1$.  The expression
in Eq.~(\ref{term1}) becomes
\begin{equation}
\bigg(\frac{1}{n-1}\bigg)\frac{\int_0^1 (1-x_1)(1-x_1)^{n-2}dx_1}
{\int_0^1 (1-x_1)^{n-2}dx_1}\int_{2,\ldots,n} f(x) dx.
\label{xxx}
\end{equation}
Here the factor of $(1-x_1)^{n-2}$ comes from the
fact that the area of the surface defined by
a fixed value of $x_1$ is proportional to
$(1-x_1)^{n-2}$.  The denominator provides the
proper normalization.
Evaluating the integrals over $x_1$ brings the
expression in Eq.~(\ref{xxx}) to
\begin{equation}
(1/n)\int_{2,\ldots,n} f(x) dx.
\end{equation}
We can treat the other terms on the right-hand
side of Eq.~(\ref{flong}) in the same way, so that
upon integration, this inequality becomes
\begin{equation}
\int f(x) dx \leq (1/n)\sum_{k_1<\cdots < k_{n-1}}
\int_{k_1,\ldots,k_{n-1}} f(x) dx.
\label{intineq}
\end{equation}
Multiplying both sides by $n$ and using
Eq.~(\ref{Rf}), 
we have
\begin{equation}
R^{(n)}_n \leq R^{(n)}_{n-1},
\end{equation}
with equality holding only if just one of the
$\lambda$'s is nonzero, that is, if $\rho$ is pure.

The other inequalities in Eq.~(\ref{ineq}) can be
obtained by a similar argument.  Consider any 
face of the probability simplex in which only
$r$ of the components $x_i$ are non-zero.  Each
point $x$ on such a face can be decomposed
as in Eq.~(\ref{long}), and the above argument
gives us an inequality analogous to Eq.~(\ref{intineq}):
\begin{equation}
\int_{k_1,\ldots,k_r} f(x)dx \leq \frac{1}{r}
\bigg[\int_{k_2,\ldots,k_r}f(x)dx
+\int_{k_1,k_3,\ldots,k_r}f(x)dx
+\cdots + \int_{k_1,\ldots,k_{r-1}}f(x)dx\bigg].
\label{dots}
\end{equation}
We now insert this inequality into the 
expression (\ref{Rf}) for $R^{(n)}_r$:
$$
R^{(n)}_r = \chose{n}{r}^{-1}n\sum_{k_1<\cdots <k_r}
\int_{k_1,\ldots,k_r} f(x) dx  
$$
\begin{equation}
\leq \chose{n}{r}^{-1}n\bigg(\frac{n-(r-1)}{r}\bigg)
\sum_{k_1<\cdots < k_{r-1}}\int_{k_1,\ldots,k_{r-1}}f(x)dx.
\label{nasty}
\end{equation}
Here the factor of $n-(r-1)$ comes from the following
fact: given any set $A$ of $r-1$ distinct index-values
[which defines the range of one of the integrals on the 
right-hand side of Eq.~(\ref{nasty})], there are 
$n-(r-1)$ sets of $r$ distinct index-values
from which $A$ could have been obtained by the 
deletion of one value, so that each integral
associated with the set $A$ appears $n-(r-1)$ times.
Simplifying the factors in Eq.~(\ref{nasty}),
we get
\begin{equation}
R^{(n)}_r \leq \chose{n}{r-1}^{-1}n
\sum_{k_1<\cdots < k_{r-1}}\int_{k_1,\ldots,k_{r-1}}f(x)dx
= R^{(n)}_{r-1}.
\end{equation}
Moreover, by an argument similar to what we used before,
equality holds only if $\rho$ is pure.
This completes our proof of the string of 
inequalities (\ref{ineq}).

\section{Other properties of $R^{(n)}_r$}

In this section we demonstrate various other properties
of $R^{(n)}_r$.  In particular: (i) we show that 
as a function of $\lambda = (\lambda_1,\ldots,
\lambda_n)$, $R^{(n)}_r$ is concave;
(ii) we find the maximum value of $R^{(n)}_r$; (iii)
we determine how $R^{(n)}_r$ is affected by 
the addition of extra dimensions with zero
eigenvalues.

\bigskip

\noindent (i) {\em $R^{(n)}_r$ is concave.}  
We showed earlier that the quantity $f$ of
Eq.~(\ref{f}), regarded as a function of $x$,
is convex.  It is easier to see that as a
function of $\lambda = (\lambda_1,\ldots,
\lambda_n)$ (with $\sum_i \lambda_i = 1$),
$f$ is {\em concave}: the function $-y\ln y$
is concave in $y$, and apart from a linear
term, our function $f$ is 
of this form, with $y$ being a linear function
of the $\lambda$'s.  According to Eq.~(\ref{Rf}),
$R^{(n)}_r$ is a sum of these concave functions
and is therefore concave itself.  

\bigskip

\noindent (ii) {\em Maximum value of $R^{(n)}_r$.}
Because $R^{(n)}_r$ is concave and is symmetric
under interchange of the $\lambda_i$'s, it must
achieve its maximum value when all the $\lambda_i$'s
are equal, in which case they are all equal
to $1/n$.  It is probably easiest to obtain this
maximum value explicitly via Eq.~(\ref{Rf}).
Upon doing the integral, one finds that
for $r=2,\ldots,n$,
\begin{equation}
\hbox{maximum of }R^{(n)}_r =
\ln n - \bigg(\frac{1}{2}+\frac{1}{3}+
\cdots + \frac{1}{r}\bigg).
\end{equation}

\bigskip

\noindent (iii) {\em Adding null dimensions.}
For many purposes, a density matrix in $n$ dimensions
can be regarded equally well as a density matrix
in $m$ dimensions with $m > n$, but with 
$m-n$ additional eigenvalues that are all zero. 
As we mentioned in the introduction, 
the entropy $S(\rho)$ does not change if one
adds dimensions in this way (just as the 
Shannon entropy does not change if one imagines
additional possibilities all having zero 
probability), and neither does the subentropy $Q$.  
It is interesting that in the case of $Q$ this 
invariance
follows immediately from the form of
Eq.~(\ref{Qint}): supplementing $\rho$ with
extra zero eigenvalues means supplementing
the matrix $(I-\rho/z)^{-1}$ with extra eigenvalues
all equal to 1, and these eigenvalues do not change the
determinant.  

As we have said, however, 
our intermediate quantities $R^{(n)}_r$ for
$r = 2, \ldots,$ \hbox{$n-1$} do not behave so simply upon 
addition of null dimensions.  From Eq.~(\ref{Rint})
one can show that adding $m$ zero eigenvalues
to what was originally an $n\times n$ density matrix
has the following effect on $R_r$:
\begin{equation}
R^{(n+m)}_r = \chose{n+m-1}{r-1}^{-1}\,
\sum_{s=0}^{r-1}\chose{n-1}{r-1-s}\chose{m}{s}
R^{(n)}_{r-s}.
\label{addzero}
\end{equation}
It is worth checking that this equation is consistent
with our assertion that both $S$ and $Q$ are invariant under the
addition of zero eigenvalues.
The entropy in $n+m$ dimensions is 
$S^{(n+m)}=R^{(n+m)}_1$.  Setting $r=1$ in the above
equation gives us just one term, the one
with $s=0$, and we see that $R^{(n+m)}_1=R^{(n)}_1$.
Similarly for the subentropy, $Q^{(n+m)} = R^{(n+m)}_{n+m}$:
if we set $r=n+m$ in the above equation, we find
again that only one term survives, the one with
$s=m$, and that $R^{(n+m)}_{n+m} = R^{(n)}_n$.

\section{Combinations invariant under the addition of 
zero eigenvalues}
Invariance under the addition
of null dimensions is a rather essential property
of the von Neumann entropy.  So
if we are looking for generalizations of entropy, we 
might reasonably insist on this invariance.  We have just
seen that $R^{(n)}_r$ with $r=2,\dots, n-1$ does not
have this property, at least not in any obvious sense,
but it is 
interesting to ask whether we can use the $R^{(n)}_r$'s
to construct
functions that {\em are} invariant in this way.  In particular,
for each value of $n$ 
let us look at {\em weighted averages} of the $R^{(n)}_r$'s.   
That is, we ask whether one
can find functions ${\mathcal R}^{(n)}(\lambda_1,\ldots,
\lambda_n)$ of the form
\begin{equation}
{\mathcal R}^{(n)} = \sum_{r=1}^n b^{(n)}_r R^{(n)}_r
\label{bnr}
\end{equation}
with $b^{(n)}_r \geq 0$ and $\sum_r b^{(n)}_r = 1$,
such that 
\begin{equation}
{\mathcal R}^{(n+1)}(\lambda_1,\ldots,\lambda_n,0)
={\mathcal R}^{(n)}(\lambda_1,\ldots,\lambda_n).
\label{invariance}
\end{equation}
We will refer to such sets of functions as
``augmentation-invariant,'' or for brevity, simply
``invariant.''  

Combining Eqs.~(\ref{bnr})
and (\ref{invariance}), we see that the condition
we want to satisfy is
\begin{equation}
\sum_{r=1}^{n+1}b^{(n+1)}_r 
R^{(n+1)}_r(\lambda_1,\ldots,\lambda_n,0)
=\sum_{r=1}^n b^{(n)}_r R^{(n)}_r(\lambda_1,\ldots,\lambda_n).
\label{RR}
\end{equation}
But according to Eq.~(\ref{addzero}) with $m=1$, 
\begin{equation}
R^{(n+1)}_r(\lambda_1,\ldots,\lambda_n,0)
=\frac{n-r+1}{n}R^{(n)}_r(\lambda_1,\ldots,\lambda_n)
+ \frac{r-1}{n}R^{(n)}_{r-1}(\lambda_1,\ldots,\lambda_n).
\end{equation}
Inserting this last relation into Eq.~(\ref{RR}) and
equating coefficients of $R^{(n)}_r$, we get the following
condition on the $b^{(n)}_r$'s:
\begin{equation}
(n-r+1)b^{(n+1)}_r+rb^{(n+1)}_{r+1}=nb^{(n)}_r.
\label{basicb}
\end{equation}
If ${\mathcal R}^{(n)}$ is to be 
augmentation-invariant, then Eq.~(\ref{basicb})
must be satisfied for all pairs $(n,r)$ such that $n\geq 1$ and
$1\leq r \leq n$. Let us say that a set $b$ of 
non-negative values $b^{(n)}_r$
is a solution to the invariance problem if it is
normalized---that is, if $\sum_r b^{(n)}_r = 1$ for 
each $n$---and if it satisfies Eq.~(\ref{basicb}).
We aim to find all such solutions.
Note that the normalization condition
$\sum_r b^{(n)}_r =1$ is 
actually guaranteed by Eq.~(\ref{basicb})
for all values of $n$ if it is true for any one value
of $n$: summing Eq.~(\ref{basicb})
over $r$ gives us $\sum_r b^{(n+1)}_r = \sum_r b^{(n)}_r$.
Notice also that the set of solutions $b$
is convex: if $b$ and $b'$ are
solutions, then $pb + (1-p)b'$ with $0\leq p \leq 1$
is also a solution.

We begin by solving a slightly different problem, in 
which we restrict the range of $n$ in Eq.~(\ref{basicb})
to $1\leq n < N$ for some integer $N$.  For this
restricted problem, we note three facts: 
(i) The solution is completely 
determined by the values of $b^{(N)}_r, r=1,\ldots,N$;
moreover {\em every} set of such values yields a solution.  
(ii) Because the set of allowed
values of the ordered set $(b^{(N)}_1,\ldots,b^{(N)}_N)$
is compact, the set of solutions to the restricted problem
is also compact.  (iii) The extreme points of the 
convex set of solutions are
generated by choosing $b^{(N)}_r = \delta_{r\hat{r}}$, with
$\hat{r}$ in the range $1\leq \hat{r} \leq N$ and $\delta$
being the Kronecker delta; that is, at the level
$n=N$, we put all the weight on one value of $r$. Any other 
normalized
set of $b^{(N)}_r$'s can be obtained as a weighted overage
of these special cases. 

Remarkably, we can write down {\em explicitly} the solution
to Eq.~(\ref{basicb}) generated by
$b^{(N)}_r = \delta_{r\hat{r}}$.   
\begin{equation}
b^{(n)}_r = \chose{n-1}{r-1}\chose{N-n}{\hat{r}-r}
\chose{N-1}{\hat{r}-1}^{-1}.
\label{bN}
\end{equation}
One can verify that these $b^{(n)}_r$'s satisfy 
Eq.~(\ref{basicb}) for $n<N$, that they are normalized, 
and that they take the values
$\delta_{r\hat{r}}$ for $n=N$.  
This solution has a simple interpretation in
basic probability theory: in a series
of $N-1$ tosses of a coin, $b^{(n)}_r$
given by Eq.~(\ref{bN}) is the probability of getting exactly
$r-1$ heads in the first $n-1$ tosses, {\em given} that in
the full set of $N-1$ tosses, the number of heads is exactly 
$\hat{r}-1$.
Again, any other solution of the restricted problem can be
obtained by taking weighted averages of the solutions
presented in Eq.~(\ref{bN}).  

We now return to the original problem, with no restriction on
the value of $n$.  As in the case of the restricted problem,
there will be a set of extreme solutions from which all other
solutions can be obtained as convex combinations.
We find these extreme solutions by taking the limit of
Eq.~(\ref{bN}) as $N\rightarrow\infty$ and 
$\hat{r}\rightarrow\infty$
while the ratio $\hat{r}/N$ approaches some value $\alpha$
in the range $0\leq \alpha\leq 1$.  This limit gives us
the following basic solutions to the invariance problem:
\begin{equation}
b^{(n)}_r = \chose{n-1}{r-1}\alpha^{r-1}(1-\alpha)^{n-r}.
\label{finalb}
\end{equation}
Again, one can verify directly that these 
$b^{(n)}_r$'s satisfy
Eq.~(\ref{basicb}).  As in the restricted problem, this
solution has a simple interpretation in terms of coin
tossing: $b^{(n)}_r$ as
given by Eq.~(\ref{finalb}) is 
the probability of getting
$r-1$ heads in $n-1$ tosses if the probability
of heads is $\alpha$.  Returning now to Eq.~(\ref{bnr})
we can identify, for each value of $\alpha$, the following
invariant set of functions ${\mathcal R}^{(n)}_\alpha$:
\begin{equation}
{\mathcal R}_\alpha^{(n)}(\rho)
\equiv\sum_{r=1}^{n}\chose{n-1}{r-1}\alpha^{r-1}
(1-\alpha)^{n-r} R^{(n)}_r(\rho).
\label{calR}
\end{equation}
That is, by taking an average over $r$ of the 
functions $R^{(n)}_r$, with the weights in the average
given by a binomial distribution, one obtains
a function that 
is invariant under the addition of null dimensions.  
Moreover, these binomial averages are the extreme
cases.  One can always generate other invariant
functions by taking convex combinations, but the binomial
averages can be regarded as the basic solutions.
To put it in other words, one can find invariant
functions by weighting the $R^{(n)}_r$'s with
{\em broader} distributions, but not with {\em narrower}
distributions.  

As $\alpha$ increases from 0 to 1, the peak of the
binomial distribution in Eq.~(\ref{calR}) moves toward larger
values of $r$.  Since we have already shown that 
$R^{(n)}_r$ decreases (or remains unchanged) as $r$
increases, we see immediately that ${\mathcal R}^{(n)}_\alpha$
is likewise non-increasing with increasing $\alpha$.  
For the extreme values $\alpha=0$ and $\alpha=1$,
we have ${\mathcal R}^{(n)}_0 = S$ and 
${\mathcal R}^{(n)}_1 = Q$.  Thus ${\mathcal R}^{(n)}_\alpha$
interpolates continuously between $S$ and $Q$.

Just as $S$ and $Q$ can be written as contour integrals,
it turns out that ${\mathcal R}^{(n)}_\alpha$ can be
written in a similar way: one can show that 
\begin{equation}
{\mathcal R}_\alpha^{(n)}(\rho)=
-\frac{1}{2\pi i\alpha}\oint(\ln z)
\det\Big\{[I-(1-\alpha)\rho/z][I-\rho/z]^{-1}\Big\} dz,
\label{calRint}
\end{equation}
where the value at $\alpha=0$ is determined by taking
the limit.   
In this form, it is quite easy to see that
${\mathcal R}^{(n)}_\alpha$ is invariant under the
addition of null dimensions.  The eigenvalues of
the matrix whose determinant we are taking in
Eq.~(\ref{calRint}) can be written as
\begin{equation}
\hbox{eigenvalues} = (1-\alpha)+\alpha\bigg(
\frac{z}{z-\lambda_i}\bigg),
\label{goodform}
\end{equation}
where as always, the $\lambda_i$'s are the eigenvalues
of $\rho$.  If any of the $\lambda_i$'s are zero, they
contribute a factor of 1 to the determinant and can 
thus be ignored in calculating the value
of ${\mathcal R}^{(n)}_\alpha$.  [The form 
(\ref{goodform}) is also particularly convenient
for deriving Eq.~(\ref{calRint}).] Because of the
augmentation-invariance, we can drop the superscript
$n$ and refer unambiguously to ${\mathcal R}_\alpha$.
We could also use the contour integral (\ref{calRint}),
which contains no explicit reference to $n$, as an
alternative definition. 

\section{Discussion}
We have identified and studied various 
functions that lie between the entropy $S$ and the
subentropy $Q$.  Our first set of such functions
$R^{(n)}_r$ emerged as a natural mathematical
generalization of Eqs.~(\ref{Sint}) and (\ref{Qint}),
and also turned out to be generalizations of
the alternative expression (\ref{Qf}) for $Q$ as an integral
over the probability simplex.  These functions share
certain properties with entropy---they are concave,
they take the value zero when all but one of the 
eigenvalues of $\rho$ are zero, and they take their
maximum value when all the eigenvalues are equal---but
unlike entropy they do not remain unchanged when one includes
additional dimensions corresponding to zero
eigenvalues of $\rho$.  

The related functions ${\mathcal R}_\alpha$
are weighted averages of the $R^{(n)}_r$'s and therefore
share the properties just listed, but in addition
they are invariant under the inclusion of null
dimensions.  Moreover they are the most basic
functions having this property: other
augmentation-invariant functions can be obtained as convex
combinations of the ${\mathcal R}_\alpha$'s.

One consequence of this invariance is a very modest
kind of additivity.  Let $\rho_1$ be an arbitrary
density matrix of some quantum system and let
$\rho_2$ be the density matrix
of a {\em pure} state of another system.  Then 
for any $\alpha$ in the range $0\leq\alpha\leq 1$,
we can say
\begin{equation}
{\mathcal R}_\alpha (\rho_1 \otimes\rho_2)
= {\mathcal R}_\alpha (\rho_1) + 
{\mathcal R}_\alpha (\rho_2).
\end{equation}
This statement follows from the augmentation-invariance of
${\mathcal R}_\alpha$ along with
two simple facts:
(i) $\rho_1$ and $\rho_1\otimes\rho_2$ have the
same {\em nonzero} eigenvalues, and (ii) 
${\mathcal R}_\alpha (\rho_2)=0$.
On the other hand, for {\em arbitrary} $\rho_1$ and
$\rho_2$, ${\mathcal R}_\alpha$ is not
additive except when $\alpha=0$, in which case
${\mathcal R}_\alpha$ is the entropy itself.

Does either $R^{(n)}_r$ or ${\mathcal R}_\alpha$
have a physical meaning?  At this point we have no 
definite interpretation of either of these quantities,
though because of its nice mathematical properties
we have more hope for ${\mathcal R}_\alpha$.
Here we suggest one way in which this quantity
might play a role in quantum information theory.

Consider once again an ensemble ${\mathcal E} =
\{(|\psi_i\rangle,p_i)\}$ of
pure states of a quantum particle, and suppose that one
is trying to convey classical information by sending
a sequence of states chosen 
from this ensemble, with frequencies
of occurrence asymptotically equal to 
the given probabilities $p_i$.  If the 
receiver (Bob) is required to measure each particle 
individually, then the maximum amount of 
information that the sender (Alice) 
can convey per particle is the accessible
information of the ensemble ${\mathcal E}$.  Suppose, 
though, that Bob is able to measure {\em pairs}
of particles jointly.  Then Alice can
hope to convey more information per particle
by encoding her message in codewords consisting
of pairs of the original states; that is,
each codeword is of the form $|\psi_{i_1}\rangle\otimes
|\psi_{i_2}\rangle$ with $|\psi_{i_1}\rangle$ and 
$|\psi_{i_2}\rangle$
chosen from ${\mathcal E}$.  We insist that Alice
respect the original probabilities of ${\mathcal E}$
in the sense that in a long message, each state
$|\psi_i\rangle$ is used with a frequency approximating
$p_i$.  One finds that Alice often ${\em can}$
increase the information conveyed per particle by using
this strategy \cite{PW,HJSWW,SKIH,BVF}.  Moreover, by continuing
to increase the length of the codewords, 
assuming that Bob can make
arbitrary joint measurements on a whole codeword, Alice
can convey even more information.  Let $I_m$ be the 
amount of information one can convey per particle when
the codeword length is $m$.  The limiting value of $I_m$ 
for arbitrarily long
codewords is simply $S(\rho)$, where $\rho$ is the density
matrix of the ensemble ${\mathcal E}$ \cite{HJSWW}.  

In the first stage of the above scenario, when Bob can
measure only individual particles, we know that $Q(\rho)$ is 
a lower bound on the information that can be
conveyed per particle.  
As the codeword length increases to infinity, 
$I_m$ increases to $S(\rho)$.  One is led to 
speculate that for intermediate codeword lengths, 
${\mathcal R}_\alpha(\rho)$ may play a role.  For
example, it is conceivable that when Alice and Bob
are using codewords of length $m$, 
${\mathcal R}_\alpha(\rho)$ is a lower bound 
on $I_m$, where $\alpha = e^{-c(m-1)}$
for some universal constant $c$.  As $m$ approaches
infinity, then, the lower bound would approach
${\mathcal R}_0(\rho) = S(\rho)$, as it should.  

We can extend this idea to the study of the 
classical capacity of a quantum channel.  At present
one does not have a simple way of calculating this
capacity for all channels, only because it is
not known whether the amount of information 
conveyed can be increased by using inputs that
are entangled between different uses of the channel
\cite{Shor}.
If we disallow entangled inputs, then the resulting
capacity---called the Holevo capacity---{\em is}
given by a simple expression \cite{SW,H}: it is the maximum,
over all input ensembles, of the quantity
$S(\rho) - \sum_i p_iS(\rho_i)$.  Here 
$\{(\rho_i,p_i)\}$ is the output ensemble, and
$\rho$ is its average density matrix 
$\sum_i p_i\rho_i$.  
As in the preceding paragraph, achieving this
capacity requires that Bob be able to make
joint measurements on arbitrarily long blocks.
But suppose that Bob cannot make such measurements;
suppose that he can measure only blocks of size $m$.
For the case $m=1$, it is known that the information
$I_1$ that he can gain per particle is bounded below
by $\max[Q(\rho)-\sum_i p_iQ(\rho_i)]$, the maximum
being over all input ensembles \cite{JRW}.
Just as
in the preceding paragraph, we can speculate that
for arbitrary $m$, the information $I_m$ that one
can convey per use of the channel is bounded
below by 
$\max[{\mathcal R}_\alpha(\rho)-\sum_i p_i
{\mathcal R}_\alpha(\rho_i)]$, with $\alpha$
given by $\alpha = e^{-c(m-1)}$.  Of course this statement
is quite speculative and we would not even want
to claim it as a conjecture.  We present it
only to suggest how the quantity ${\mathcal R}_\alpha$
might conceivably be applied.  

What we do have at present are a set of 
functions that share some mathematical properties
with entropy and subentropy.  There is a certain
elegance in the mathematics, but whether this 
elegance translates into value for physics remains
to be seen.

\end{document}